# Crystal-Symmetry-Dependent Field-Free Superconducting Diode Effect in CoPt/Nb Heterostructures

He Yu,‡ Jiaxin Chen,‡ Yi Zhang, Gan Zhao, Yizhen Shao, Jiaxu Li, Yu He, Qiusha Wang, Yuanliang Zhu, Bin Han, Junda Qu, Haochang Lyu, Bowen Dong, Jie Qi, Dapeng Zhu, Daoqian Zhu, Yunchi Zhao, Liang Liu,* Shouguo Wang,* Weisheng Zhao,* and Guang Yang*

**ABSTRACT**: Superconducting diodes convert nonreciprocal critical currents into current-polarity-selective switching between zero-resistance and resistive states. However, realizing the superconducting diode effect often requires complex device architectures and an external magnetic field to break spatial-inversion and time-reversal symmetry, substantially limiting its integration into cryogenic electronics and superconducting quantum circuits. Here, we demonstrate precise control of the field-free superconducting diode effect in epitaxial [111]-oriented A1-CoPt/Nb heterostructures. The critical-current asymmetry obeys the magnetic-point-group symmetry relation: it is finite along low-symmetry directions, suppressed along magnetic mirror axes, and reverses under either a 60° in-plane rotation or reversal of the CoPt magnetization. The same threefold symmetry dependence is observed in field-free current-induced magnetization switching in A1-CoPt films, indicating a common crystallographic origin for the superconducting and normal-state transport responses. Current- and temperature-dependent second-harmonic transport measurements are consistent with vortex-mediated nonreciprocity arising from an orientation-dependent interfacial edge landscape. We further exploit the magnetization-dependent diode polarity for proof-of-concept superconducting readout of electrically written magnetic states. These results establish crystallographic symmetry as an effective degree of freedom for engineering the superconducting diode effect, expanding the accessible materials space and providing a practical route toward field-free superconducting electronics.

# INTRODUCTION

Nonreciprocal superconducting transport converts current reversal into inequivalent superconducting responses. Its most direct manifestation is the superconducting diode effect (SDE)[1–7], in which the positive and negative critical currents are unequal, $I_C^+ \neq |I_C^-|$. The signed critical-current asymmetry, $\Delta I_C = I_C^+ - |I_C^-|$, defines the polarity of the diode response. Within the resulting diode window, a superconducting channel can remain in a zero-resistance state for one current polarity while switching into a resistive state for the opposite polarity. This current-direction-selective transition has motivated broad interest in superconducting rectifiers[8,9], cryogenic logic[10,11], and quantum devices[12–14].

For reconfigurable superconducting electronics, however, a fixed-polarity diode is only a first step. A more powerful functionality is obtained when the sign of $\Delta I_C$ can be reversed at zero applied magnetic field. In that case, superconducting nonreciprocity becomes a programmable response rather than a static property set during device fabrication[10–12,15]. A magnetic degree of freedom provides a natural nonvolatile control variable: if switching the magnetization reverses $\Delta I_C$, a single read current chosen inside the diode window can place the device in either a superconducting or a resistive state depending on the magnetic state[2,10–12,15,16]. Magnetically switchable field-free superconducting diodes therefore offer a route to encode magnetic information directly into superconducting transport. Realizing such operation requires control over two independent symmetry ingredients. The first is a time-reversal-symmetry-breaking (TRSB) component, which can be supplied by an external magnetic field[1,17], magnetic proximity[18–23] or magnetic stray fields[24,25]. The second is a spatial inversion-symmetry-breaking (ISB) axis that distinguishes opposite current directions and fixes the sign of the diode response for a given magnetic state. In many SDE platforms, this spatial axis is introduced extrinsically through asymmetric device geometries[8,17,24,26–28], van der Waals stacks[3,11,12] or multilayer superlattices[1,2]. These approaches have enabled important progress, including field-free and magnetically switchable superconducting diodes. However, they often struggle to combine zero-field operation with simple film stacks and device architectures[10–12,16], limiting their scalability. A central question is whether this spatial axis can instead be selected by an intrinsic crystalline degree of freedom.

Vortex-mediated superconducting nonreciprocity provides a useful physical setting in which this question can be made concrete. In type-II superconducting strips, the onset of dissipation can be governed by vortex entry over surface or edge barriers, and unequal barriers at opposite edges are known to generate nonreciprocal critical currents[8,17,24,27,29]. Ferromagnet/superconductor hybrids further show that magnetic textures, stray fields and screening currents can steer vortex motion and control the sign of the vortex diode response[24,30]. These studies establish asymmetric vortex dynamics as a physically grounded route to SDEs. They also indicate that vortex-mediated transport is highly sensitive to signed edge or interfacial asymmetries, making it a suitable probe of whether a microscopic crystallographic symmetry can imprint its polarity onto superconducting nonreciprocity.

A1-phase $Co_{30}Pt_{70}$ (CoPt) brings together the two symmetry ingredients needed for this test[31,32]. Its perpendicular magnetic anisotropy (PMA) stabilizes a reversible out-of-plane magnetization, providing a magnetization-dependent TRSB component for the adjacent Nb layer. At the same time, the [111]-projected $C_{3v}$ lattice contains threefold rotational symmetry and mirror-symmetric in-plane axes, offering crystallographic directions that can either relate or distinguish the transverse edges of a patterned Hall bar. Using nominally identical A1-CoPt/Nb devices patterned along different crystal orientations, we find that the zero-field SDE is selected by the A1-CoPt lattice. The signed diode response is allowed along low-symmetry directions, nearly quenched along mirror-symmetric axes and sign-reversed by a 60° crystal rotation. This angular pattern is absent in Nb control devices but follows the field-free spin-orbit-torque (SOT) response of CoPt, identifying the low-symmetry lattice as the symmetry source. Second-harmonic transport and magnetization-dependent superconducting transport measurements are consistent with a vortex-mediated mechanism in which CoPt crystal orientation biases the effective vortex-entry landscape, whereas its perpendicular magnetization switches the diode polarity. We further use this magnetization-dependent polarity for proof-of-concept superconducting readout of electrically written CoPt magnetization states. Our results establish low-symmetry magnetic crystals as intrinsic symmetry elements for designing polarity-programmable field-free superconducting diodes.

# RESULTS AND DISCUSSION

## Characterization of A1-CoPt films with $C_{3v}$ symmetry

To investigate the role of crystal symmetry in superconducting nonreciprocal transport, we first deposited a 6-nm-thick A1-CoPt film on a MgO (111) single-crystal substrate and characterized its structural and magnetic properties. The X-ray diffraction (XRD) $\theta - 2\theta\ scan$ reveals a pronounced CoPt diffraction peak at 41.1°, confirming the high crystalline quality of the film **(Figure S1a, Supporting Information**). The XRD $\varphi - scan$ further exhibits a threefold rotational symmetry, consistent with the $C_{3v}$ symmetry of A1-CoPt (**Figure S1b, Supporting Information**). **Figure 1c** presents the anomalous Hall response of the CoPt film. The nearly square hysteresis loop and a remanence ratio close to $100\%$ indicate robust PMA. This out-of-plane magnetization generates a stray field in the adjacent superconducting layer, thereby breaking TRS.

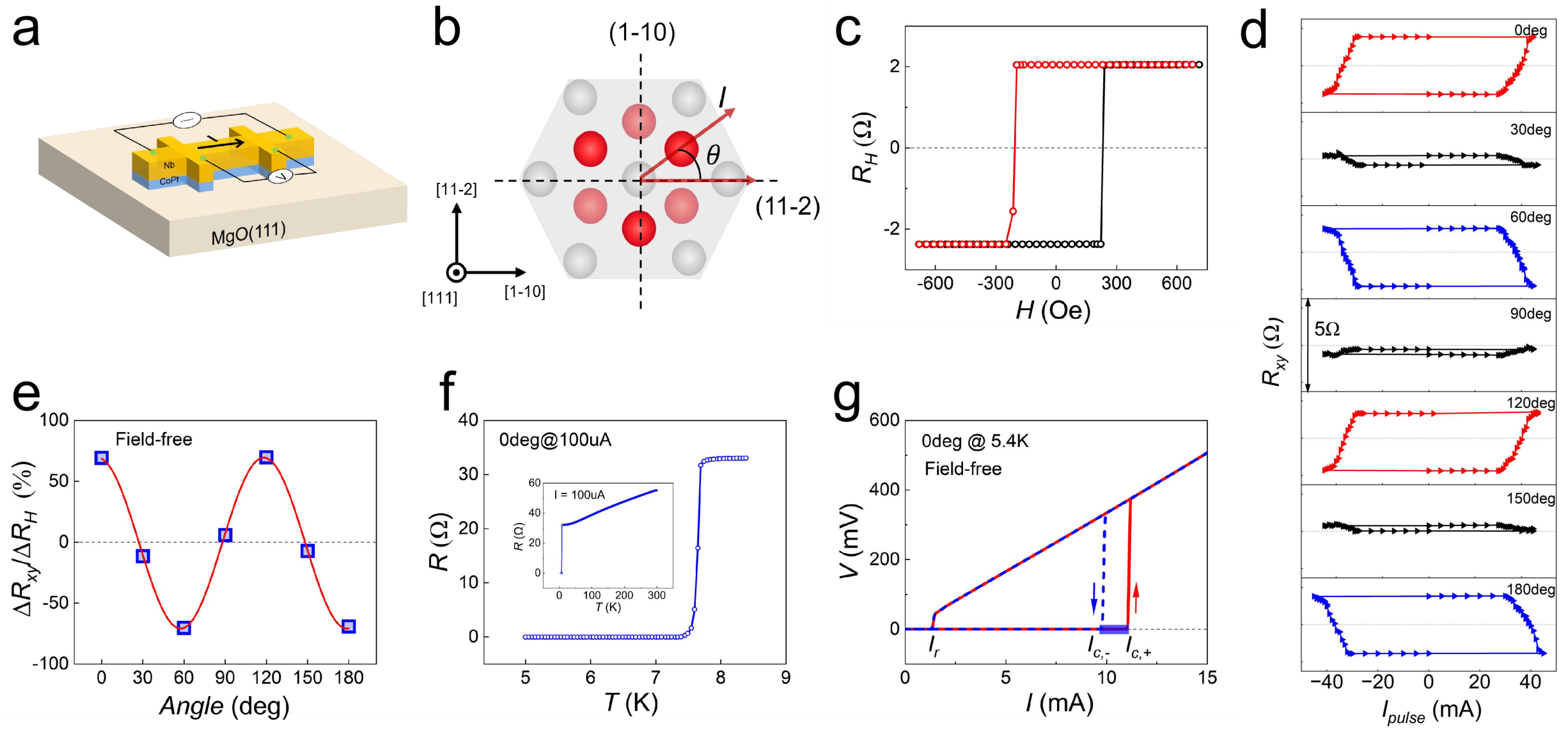

**Figure 1. Crystal symmetry, current-induced magnetization switching and superconducting transport characterization.** a) Schematic of the CoPt/Nb heterostructure and Hall-bar device geometry. b) Schematic local Co-platelet/Pt interfacial motif in A1-CoPt along the [111] direction. Co and Pt atoms are represented by red and gray spheres, respectively. The Pt layer is sandwiched between two Co layers, with the upper and lower Co sublayers shown in dark and light red. The structure corresponds to an ABC-stacked face-centered cubic (fcc) lattice along [111]. For clarity, the CoPt [1-10] direction is defined as the x-axis, and $\theta$ denotes the angle between the current direction and x-axis. c) Anomalous Hall effect of the A1-CoPt film. d) Current-induced SOT magnetization

switching under zero magnetic field for devices with different crystallographic orientations $\theta = 0°, 30°, 60°, 90°, 120°, 150°$ and $180°$. e) Normalized switching efficiency $\Delta R_{xy}/\Delta R_H$ extracted from (d) as a function of $\theta$. f) Temperature dependence of the resistance $(R - T)$ of a $\theta = 0°$ CoPt/Nb device measured under a dc current of 100 μA. The inset shows the cooling curve from 300 K to 5 K. To minimize thermal hysteresis effects during the superconducting transition, the main panel presents the warming curve from 5 K to 8.5 K. g) Field-free $I - V$ characteristics of the $\theta = 0°$ device at $0.7\,T_C$. The negative current branch is plotted in absolute value for clarity.

To probe the transport consequences of the CoPt crystal symmetry, we next examined current-induced magnetization switching in devices patterned along different crystallographic directions. As shown in **Figure 1d**, robust field-free switching is observed for the current applied along $\theta = 0°$, $60°$, $120°$ and $180°$, whereas the switching signal is strongly reduced along the mirror symmetry direction for $\theta = 30°$, $90°$ and $150°$. Additionally, the angular dependence of the switching efficiency $(\Delta R_{xy}/\Delta R_H)$ is summarized in **Figure 1e**, which exhibits a clear threefold periodicity that closely follows the $C_{3v}$ symmetry of CoPt. These measurements confirm that the intrinsic crystal symmetry imposes a pronounced directional selectivity on the current-induced magnetization switching.

We then deposited a 30-nm-thick polycrystalline Nb layer on A1-CoPt film to form **MgO(111)//CoPt(6 nm)/Nb(30 nm)** heterostructure and patterned it into Hall-bar devices with a width of 5 um (**Figure 1a**). The temperature-dependent resistance $(R - T)$ is shown in **Figure 1f**, from which a superconducting transition temperature $(T_C)$ of 7.65 K is obtained using the criterion that the resistance drops to 50% of its normal-state value, indicating a well-defined superconducting transition in the film. We subsequently measured the current–voltage $(I - V)$ characteristics at zero applied magnetic field to determine whether the heterostructure supports nonreciprocal superconducting transport. **Figure 1g** shows representative $I - V$ curves recorded at $0.7\,T_C$ with the current applied along the angle $\theta = 0°$. The critical currents, $I_C^+$ and $I_C^-$ correspond to the transitions from the superconducting state to the resistive state under positive ($-I$ to $+I$) and negative ($+I$ to $-I$) current bias, respectively. A clear asymmetry is observed, $I_C^+ \neq |I_C^-|$, demonstrating a pronounced SDE. By contrast, the retrapping currents $(I_r)$ are nearly identical for the two current polarities. This distinction indicates that the nonreciprocity originates primarily from the instability of the superconducting state rather than from asymmetric thermal retrapping.

## Angle-dependent field-free superconducting diode effect

To further elucidate the origin of the field-free SDE, we systematically investigated the superconducting transport response of devices patterned along different crystallographic orientations. Prior to the transport measurements, the film was initialized into a well-defined $+M$ state by applying a $+1$ T out-of-plane magnetic field at 10 K, which is well above its coercive field of approximately $5100$ Oe (**Figure S2, Supporting Information**). As shown in **Figure 2a-g**, the out-of-plane magnetic field was swept from $+40$ Oe to $-40$ Oe, and the field-dependent critical currents were subsequently measured at $0.7\,T_C$. The positive and negative critical currents, $I_C^+$ and $|I_C^-|$, were then extracted and tracked as functions of $H$. For devices oriented at $\theta = 0°$, $60°$, $120°$ and $180°$, the two critical-current branches are distinctly separated at the measured field range, revealing pronounced nonreciprocal superconducting transport. In sharp contrast, $I_C^+$ and $|I_C^-|$ nearly coincide for $\theta = 30°$, $90°$ and $150°$, indicating that the SDE is strongly suppressed along these crystallographic directions.

To quantify the strength of the nonreciprocal response, we define $\Delta I_C = I_C^+ - |I_C^-|$. **Figure 2h** shows the resulting $\Delta I_C - H$ curves for all device orientations. For $\theta = 0°$, $60°$, $120°$ and $180°$, $\Delta I_C$ is approximately antisymmetric with respect to the applied magnetic field, which is a characteristic signature of SDE[1,10,12,15,17,18,23,24,33]. Importantly, the polarity of the diode response for $\theta = 0°$ and $120°$ devices is opposite to that observed at $\theta = 60°$ and $180°$. This sign reversal shows that the direction of the nonreciprocal transport is set by the crystallographic orientation. In contrast, $\Delta I_C$ remains close to zero for $\theta = 30°$, $90°$ and $150°$, confirming that the SDE is strongly suppressed along these directions. We further extracted $\Delta I_C$, and the corresponding diode efficiency $\eta = \frac{I_C^+ - |I_C^-|}{I_C^+ + |I_C^-|}$ at zero-field for all device orientations. As summarized in **Figure 2i, j**, both quantities display a pronounced threefold angular dependence, which closely matches that of the field-free SOT switching shown in **Figure 1e**, pointing to a possible common origin rooted in the crystal symmetry of A1-CoPt.

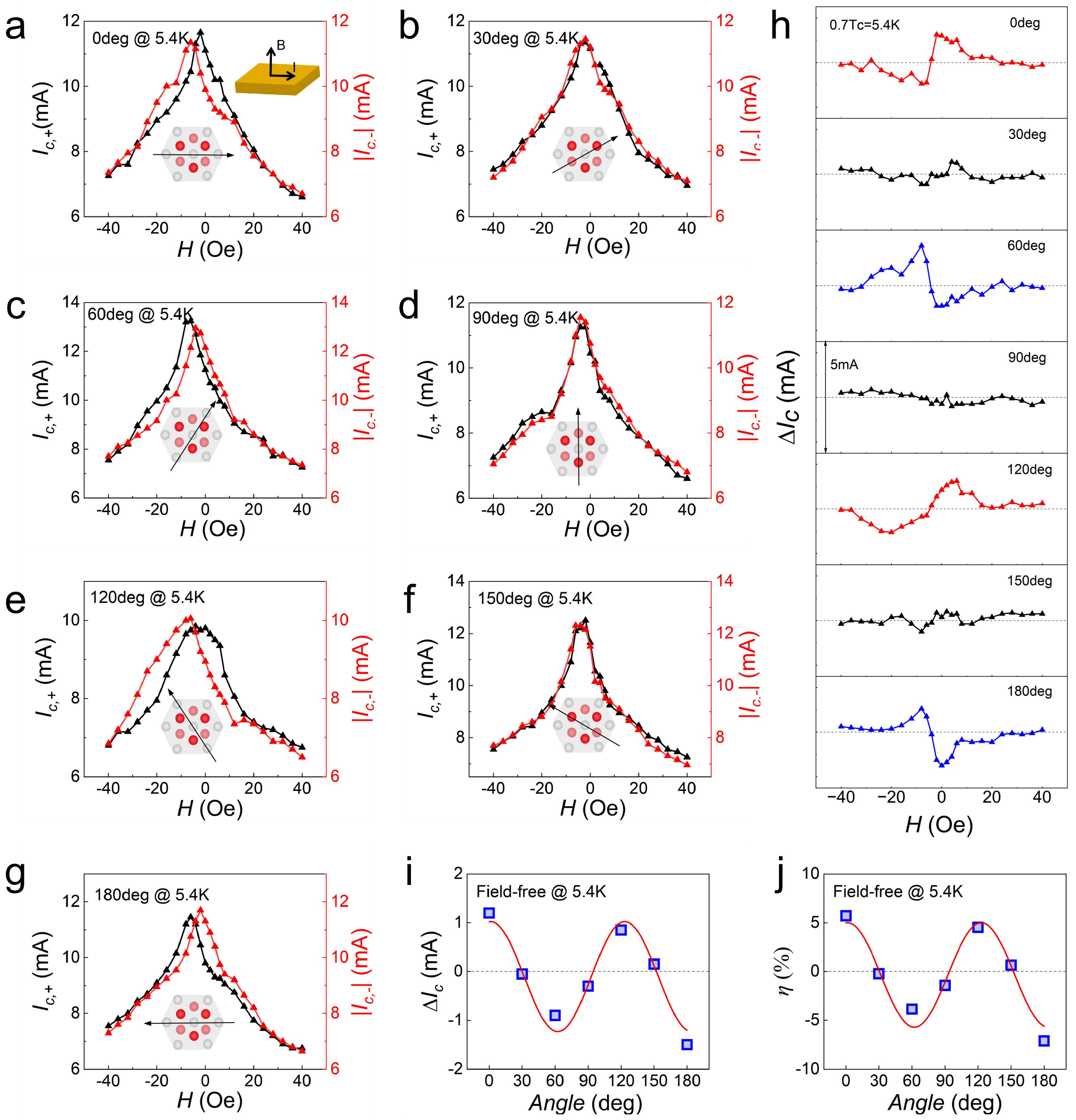

**Figure 2.** Angle-dependent superconducting diode effect in CoPt/Nb heterostructures. a-g) Field-dependent critical-current measurements for CoPt/Nb devices patterned along different crystallographic orientations, corresponding to θ = 0° , 30° , 60° , 90° , 120° , 150°and 180°. Measurements were performed at 0.7 $T_C$ = 5.4 K while sweeping the out-of-plane magnetic field from 40 Oe to −40 Oe. At each field value, an I − V curve was recorded to extract the positive and negative critical currents $I_C^+$ and $I_C^-$. For clarity, $I_C^-$ is plotted in absolute value, allowing direct comparison between $I_C^+$ and $|I_C^-|$ as a function of H . h) Field-dependent $\Delta I_C$ for devices with different crystallographic orientations, demonstrating the strong orientation dependence of the SDE. The scale bar corresponds to 5 mA. i) Zero-field critical-current asymmetry $\Delta I_C = I_C^+ - |I_C^-|$ extracted from a-g) as a function of θ.

A clear threefold angular dependence is observed, reflecting the intrinsic crystal symmetry of CoPt. j) Zero-field SDE efficiency η extracted from a-g) as a function of θ, showing the same threefold symmetry.

The pronounced angular dependence of the SDE reveals a direct connection between superconducting nonreciprocity and the crystal symmetry of CoPt. The SDE is observed only for specific orientations, whereas it is strongly suppressed along the mirror-symmetric axes. Such a crystallographic selection rule cannot be accounted for by the device geometry, which is nominally identical for all measured devices. Instead, it demonstrates that the emergence and polarity of the SDE are governed by the intrinsic $C_{3v}$ symmetry of CoPt. We further fabricated **MgO(111)//CoPt(8 nm)/Nb(30 nm)** heterostructure and repeated the same transport measurements. These devices show the same crystallographic selection rule and the same threefold angular dependence of SDE (**Figure S8, S9, Supporting Information**), which confirms that the symmetry-controlled SDE is robust and reproducible. To exclude possible contributions from the Nb layer or device geometry, we deposited a 30 nm Nb film on a single-crystalline Si substrate and patterned it into devices with the same geometry. The devices were characterized using the same $I - V$ measurement protocol, and the corresponding $\Delta I_C - H$ curves were obtained for different device orientations. In contrast to the CoPt/Nb heterostructures, $\Delta I_C$ remains nearly zero throughout the measured magnetic-field range for all orientations (**Figure S12, Supporting Information**). This control experiment shows that neither the Nb layer nor the nominal device geometry can account for the observed angular-dependent SDE.

Overall, these results show that the crystal symmetry of CoPt governs not only the angle-dependent field-free SOT switching, but also the emergence and polarity of the SDE. They therefore provide direct experimental evidence that intrinsic crystal symmetry can be used to engineer nonreciprocal superconducting transport.

## Vortex-mediated superconducting nonreciprocity

Although the measurements above establish a clear link between the SDE and the crystal symmetry of CoPt, the microscopic origin of the nonreciprocal response remains to be identified. In general, superconducting nonreciprocity can arise from two broad classes of mechanisms: an intrinsic asymmetry of the superconducting pairing state induced by spin–orbit coupling[18,23,34], or asymmetric dissipation associated with vortex dynamics[14,17,24,29,30]. To distinguish between these possibilities, we first performed second-harmonic transport measurements on the CoPt/Nb heterostructures, which can separate the linear first-harmonic voltage response ($V^{\omega}$) from the nonlinear second-harmonic component ($V^{2\omega}$), while the nonreciprocal coefficient $\gamma = \frac{\sqrt{2}V^{2\omega}}{V^{\omega}IB}$ provides a quantitative measure of the strength of the nonreciprocal transport[1,17,35].

**Figure 3a, b** show the first- and second-harmonic voltage responses measured at different excitation currents ($f = 133\ \mathrm{Hz}$). At low current, the second-harmonic signal is barely detectable. As the current increases, it rises rapidly, reaches a maximum near $2\ \mathrm{mA}$, and then decreases at higher bias. The extracted nonreciprocal coefficient $\gamma$ displays the same non-monotonic current dependence. It remains close to zero in the low-current regime, increases sharply to a maximum value of $1070\ \mathrm{T}^{-1}\mathrm{A}^{-1}$, and subsequently decreases as the current is further increased. The behavior is consistent with a vortex-dynamics mechanism[17,36–39]. At low current, the Lorentz force ($F_L = J \times \phi_0$) is insufficient to overcome the vortex-entry barrier and drive appreciable vortex motion, resulting in an almost reciprocal response. At intermediate current, vortices become mobile and their direction-dependent motion produces pronounced nonreciprocal dissipation. At still higher current, Joule heating and current-induced suppression of the superconducting order parameter become increasingly important, thereby reducing the relative nonreciprocal response.

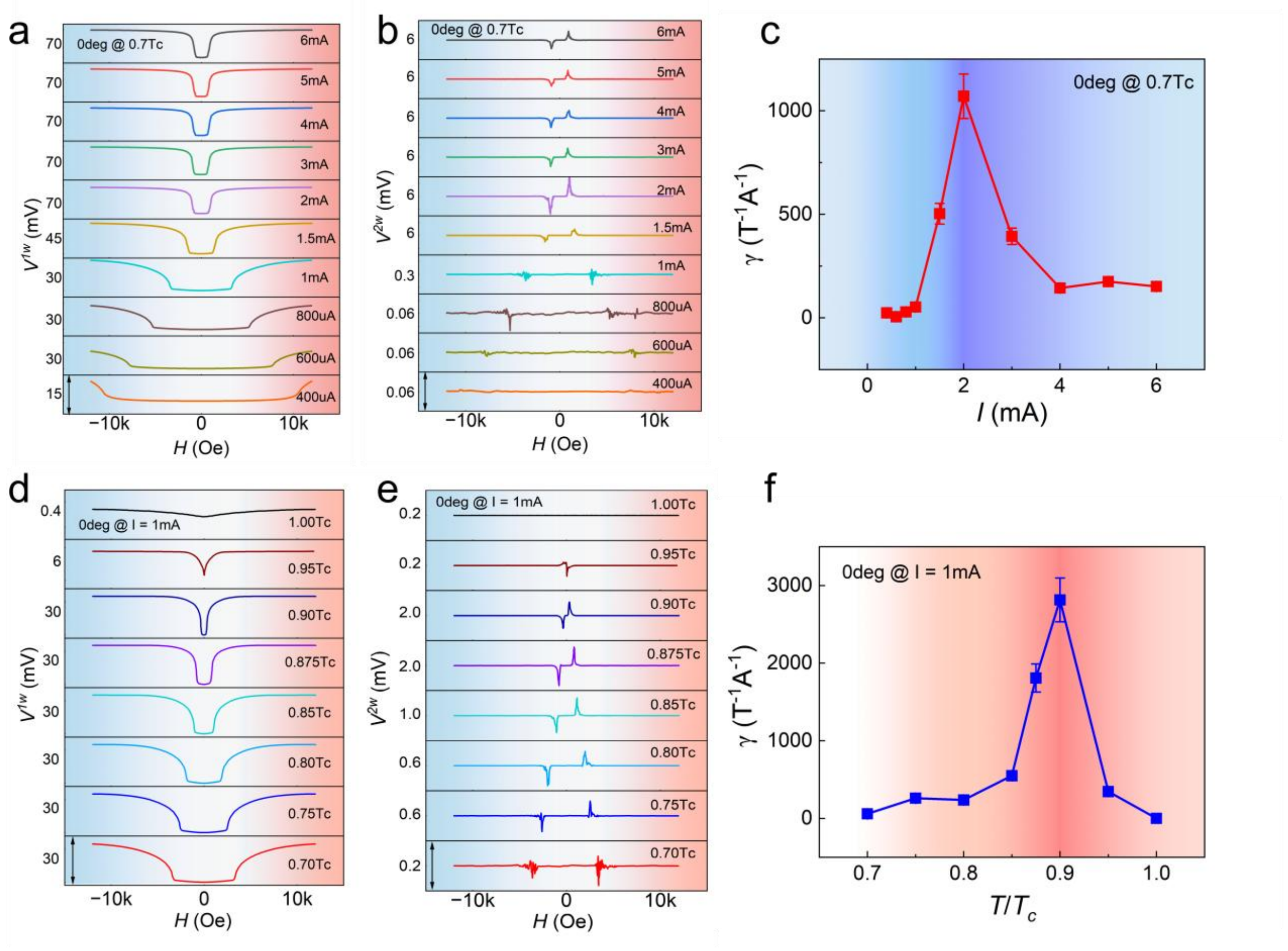

**Figure 3.** Current- and temperature-dependent superconducting nonreciprocal transport in CoPt/Nb heterostructures. a, b) First-harmonic ($V^{\omega}$) and second-harmonic ($V^{2\omega}$) voltage responses of a $\theta = 0°$ CoPt/Nb device measured at $5.4$ K under different excitation currents. An out-of-plane magnetic field is swept in a full loop from $+1.2$ T to $-1.2$ T and back to $+1.2$ T with a step size of $20$ Oe. To clearly resolve the harmonic responses under different excitation currents, different voltage scales are used as indicated in the corresponding axes. c) Current dependence of the nonreciprocal coefficient $\gamma$ extracted from a) and b). $V^{2\omega}$ is taken from the characteristic peak of the second-harmonic signal, while $V^{\omega}$ is evaluated at the corresponding magnetic field position of this peak. Error bars arise from uncertainties in peak position and signal extraction due to the finite magnetic-field step size. d, e) First- and second-harmonic voltage responses of a $\theta = 0°$ device at different reduced temperatures $T/T_C$. The excitation current is fixed at $1$ mA, and the measurement protocol is identical to that used in a) and b). f) Temperature dependence of $\gamma$ extracted from d) and e), following the same analysis procedure as in c). The data extraction procedure and the origin of the error bars are the same as those used in c).

We next examined the temperature dependence of the nonreciprocal response, as shown in **Figure 3d, f**. At low temperatures ($0.7\,T_C - 0.85\,T_C$), $\gamma$ remains small, then increases progressively with increasing temperature and reaches a maximum ($2815\ \mathrm{T^{-1}A^{-1}}$) near $0.9\,T_C$.

As the temperature approaches $T_C$, $\gamma$ rapidly decreases and eventually vanishes. This evolution reflects the temperature-dependent dynamics of superconducting vortices[17,36–39]. At low temperatures, vortices are strongly pinned and their motion is largely suppressed, resulting in only weak nonreciprocal dissipation. With increasing temperature, thermal activation lowers the effective pinning barrier and enhances vortex mobility, leading to a stronger nonlinear response. Closer to $T_C$, however, thermal fluctuations weaken the superconducting condensate and suppress the well-defined vortex state, causing the nonreciprocal signal to collapse. Taken together, the current- and temperature-dependent second-harmonic measurements support a vortex-dynamics-dominated origin of the superconducting nonreciprocity.

## Symmetry analysis and phenomenological vortex-barrier picture

The results above indicate that crystal symmetry can modulate superconducting vortex dynamics in CoPt/Nb heterostructures. However, the microscopic link between crystal symmetry and asymmetric vortex barriers remains unclear. To account for this symmetry dependence, we consider an effective vortex-dynamics model in which the interfacial $C_{3v}$ symmetry allows orientation-dependent superconducting boundary conditions and, consequently, inequivalent vortex-entry barriers in the Nb layer (**Figure 4a**). Within the Ginzburg–Landau framework, the superconducting order parameter $\psi(r)$ is determined not only by the bulk GL equations, but also by the boundary conditions imposed at the interface[40–42]. Importantly, the CoPt-induced effect does not need to penetrate the entire 30-nm-thick Nb layer. Rather, the crystallographic-orientation-dependent boundary condition at the CoPt/Nb interface reshapes $\psi(r)$ in the finite-thickness Nb film, thereby leading to inequivalent Bean–Livingston barriers for vortex entry and motion.

As illustrated in **Figure 4b**, when the current is injected along a low-symmetry crystallographic direction ($\theta = 0°$), the two transverse edges of the device are no longer symmetry-equivalent. The superconducting boundary conditions at the two edges therefore become inequivalent, leading to an asymmetric distribution of the vortex barriers. For clarity, we consider the case in which the upper edge has a higher barrier than the lower edge, which can be equivalently expressed as an additional critical-current-density offset ($\delta j_C$) at the upper edge. Under out-of-plane magnetic field ($B_z$), a Meissner screening current density ($j(B_z) = \alpha B_z$, $\alpha$ is a constant) is generated with opposite directions at the two edges of the superconductor. The local current density at each edge is determined by the superposition of the applied transport current and the Meissner screening current[24,29], which can be expressed as $j_{ext} \pm \alpha B_z$. Once the local current density exceeds the edge-dependent critical value ( $j_C$ or $j_C + \delta j_C$ ), vortices overcome the vortex barrier and move under the Lorentz force. Specifically, for the upper edge, $j_{ext}^{+} + \alpha B_z = j_C + \delta j_C$; whereas for the lower edge, $j_{ext}^{-} + \alpha B_z = j_C$. Thus, $\Delta j_C = j_{ext}^{+} - j_{ext}^{-} = \delta j_C$.

By contrast, when the current is applied along a mirror-symmetric crystallographic direction ($\theta = 30°$) (**Figure 4c**), the two edges remain symmetry-equivalent. In this case, the barriers at the two edges are the same. Vortex entry and motion are therefore symmetric for opposite current directions, resulting in $j_{ext}^{+} = j_{ext}^{-}$, which explains the disappearance of the SDE for $\theta = 30°, 90°$ and $150°$.

For $\theta = 60°$ direction (**Figure 4d**), the two transverse edges are again not symmetry-equivalent. However, because the crystallographic asymmetry is reversed by $180°$ relative to $\theta = 0°$ direction, the edge-barrier asymmetry is reversed as well. As a result, for the upper edge, $j_{ext}^{+} + \alpha B_z = j_C$; whereas for the lower edge, $j_{ext}^{-} + \alpha B_z = j_C + \delta j_C$. Thus, $\Delta j_C = j_{ext}^{+} - j_{ext}^{-} = -\delta j_C$, leading to a reversal of the SDE polarity. This accounts for the opposite signs of the diode response observed between $\theta = 60°$, $180°$ devices and $\theta = 0°$, $120°$ devices in **Figure 2j**.

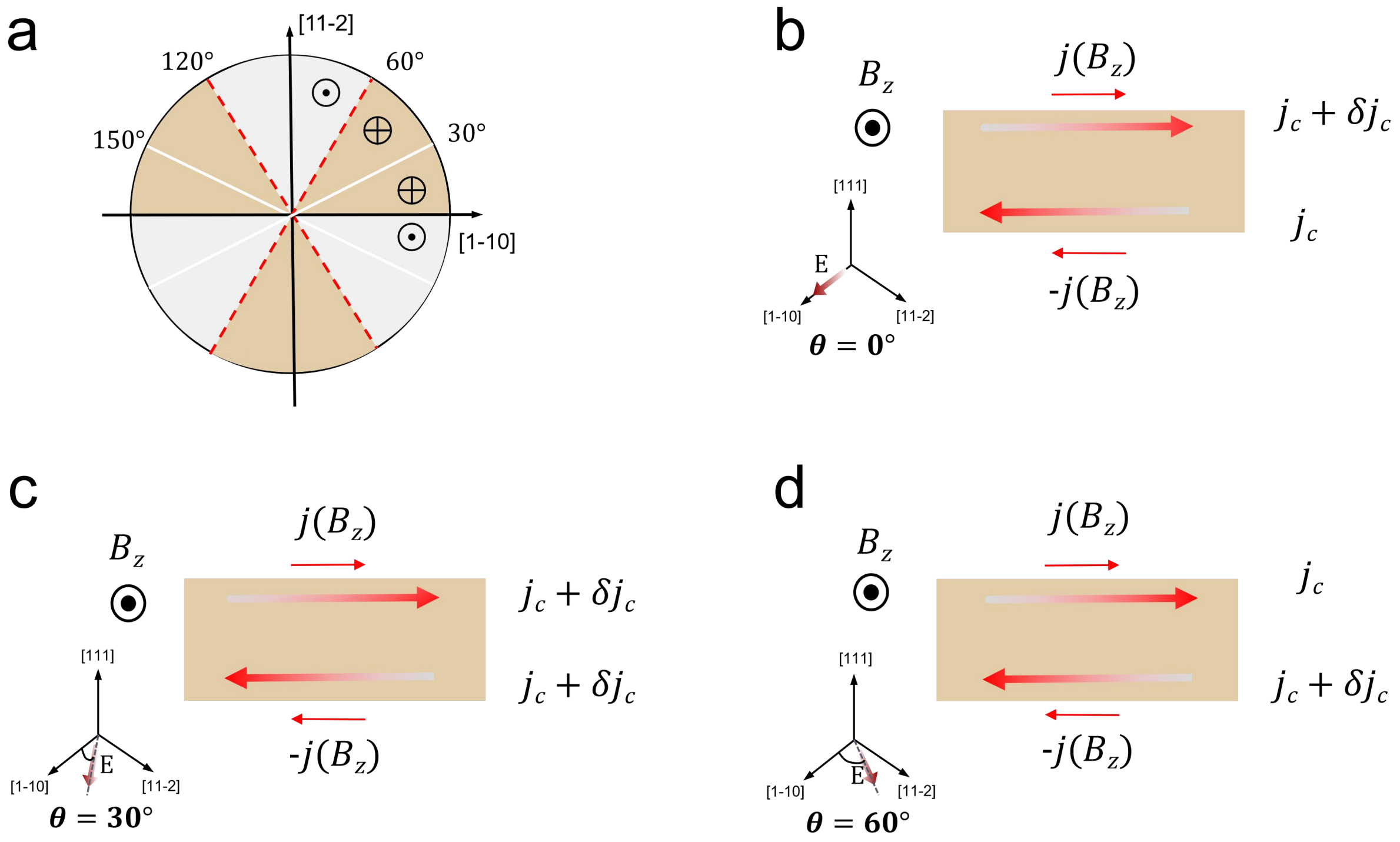

**Figure 4.** Phenomenological model of crystal-symmetry-controlled asymmetric vortex dynamics. a) Schematic of the crystal-field symmetry at the CoPt/Nb interface. Orange and gray regions represent two distinct local crystal-field environments originating from the A1-CoPt lattice. Along the [1-10] direction, the two device edges experience inequivalent crystal-field environments. Along $\theta = 30°$, the two edges are symmetry-equivalent, preserving mirror symmetry. Along $\theta = 60°$, the edge environments are again inequivalent but with reversed asymmetry compared to the [1-10] direction, leading to a reversal of the IS breaking polarity. b) For current injection along the low-symmetry direction ($\theta = 0°$). c) For $\theta = 30°$, the two edges remain symmetry-equivalent. d) For $\theta = 60°$, the asymmetry of the vortex-entry barriers is reversed due to the rotated crystal symmetry.

The vortex-dynamic model further yields a direct and testable prediction. In CoPt with PMA, the local stray field is determined by the out-of-plane magnetization of CoPt[24]. Switching the magnetization therefore reverses the local stray field and the Lorentz force acting on the vortices. If the field-free SDE is governed by asymmetric vortex motion, the diode polarity should be reversed when the CoPt magnetization is switched[30]. For $\theta = 0°$ device, we further applied a $-1$ T out-of-plane magnetic field at 10 K to initialize the CoPt layer into the $-M$ saturated magnetization state. We then performed field-dependent $I - V$ measurements at 0.7 $T_C$ and extracted the corresponding $\Delta I_C - H$ relation (**Figure S5, Supporting Information**). The two magnetization states give rise to approximately antisymmetric nonreciprocal signals over the entire measured field range, which can be given by $\Delta I_C(+H, +M) = -\Delta I_C(-H, -M)$. In particular, the sign of $\Delta I_C$ switches when the CoPt magnetization is reversed at zero field, which can be expressed by $\Delta I_C(+M) =- \Delta I_C(-M)$. This magnetization-controlled polarity reversal confirms the time-reversal-odd nature of the diode response and is consistent with the proposed magnetically controlled vortex-dynamics scenario. Notably, the same result is observed in **MgO(111)//CoPt(8 nm)/Nb(30 nm)** heterostructure (**Figure S10, Supporting Information**), confirming the robustness and reproducibility of this phenomenon.

The angular selection rule of this vortex-barrier asymmetry can also be constrained by the **magnetic point group** of the perpendicularly magnetized CoPt layer. The nonmagnetic [111]-projected A1-CoPt lattice has the crystallographic point group $C_{3v}$. Once a single out-of-plane magnetization $M_z$ is selected, the relevant symmetry becomes the magnetic point group of the magnetized single domain. Because $M_z$ is an axial vector, a vertical mirror operation changes its sign and is therefore not by itself a symmetry operation of a fixed magnetic domain. The mirror operation survives only when combined with time reversal, giving the magnetic point group

$$G_M = \{E, C_{3z}^{+}, C_{3z}^{-}, \mathcal{T}m_1, \mathcal{T}m_2, \mathcal{T}m_3\} \equiv 3m'.$$

This magnetic point group imposes a simple selection rule on the signed critical-current asymmetry. The threefold rotations require the same response after a $120°$rotation, whereas the anti-unitary mirror-time operations $\mathcal{T}m_i$ relate positive- and negative-current critical processes along the magnetic mirror axes. As a result, the leading symmetry-allowed form of the field-free critical-current asymmetry is

$$\boldsymbol{\Delta I_c(\theta,\ M_z) \propto M_z \cos 3\,\theta}.$$

This form vanishes along the magnetic mirror axes $\theta = 30°$, $90°$ and $150°$, changes sign under a $60°$ rotation, and reverses when $M_z$ is switched. A full derivation is provided in **Note 6, Supporting Information**.

Taken together, these results show that CoPt provides two essential prerequisites for the field-free SDE in this heterostructure. First, the $C_{3v}$ symmetry of A1-CoPt imposes crystallographic-orientation-dependent superconducting boundary conditions at the CoPt/Nb interface, which modulate the Bean–Livingston vortex barriers in the Nb layer. Second, the PMA of CoPt generates a local stray field that controls the direction of vortex motion and breaks TRS.

The cooperation of these two effects gives rise to a field-free SDE that is both crystal-symmetry-dependent and magnetic-state-controllable.

## Electrical writing and superconducting readout of magnetic states

**Figure 5a** illustrates the design of an integrated write–read architecture based on the CoPt/Nb heterostructures, in which electrical writing and superconducting readout are implemented within a single device platform. At room temperature, large current pulses are used to switch the CoPt magnetization between $+M$ and $-M$. At superconducting temperature, a read current, much smaller than the magnetization-switching threshold, is applied to probe the superconducting diode response. Because the polarity of the field-free SDE is determined by the magnetic state of CoPt, the same read current can place the device either in the dissipationless superconducting state or in a finite-resistance state[10–12]. In this way, the magnetization state is converted into a binary electrical output, while the write and read operations are separated by their current scales.

We first verified that the CoPt/Nb device retains field-free current-induced magnetization switching after integration with the 30-nm-thick Nb layer. The inset of **Figure 5b** shows the anomalous Hall hysteresis loop measured at $300$ K, demonstrating that the heterostructure still maintains excellent PMA. Owing to the parallel shunting through the Nb layer, the Hall-resistance amplitude is reduced compared with that of CoPt. Nevertheless, the two perpendicular magnetization states ($+M$ and $-M$) remain clearly distinguishable. Under zero external magnetic field, positive and negative current pulses with a duration of $2$ ms switch the device reproducibly between two stable Hall-resistance states, confirming reliable all-electrical writing of the CoPt magnetization.

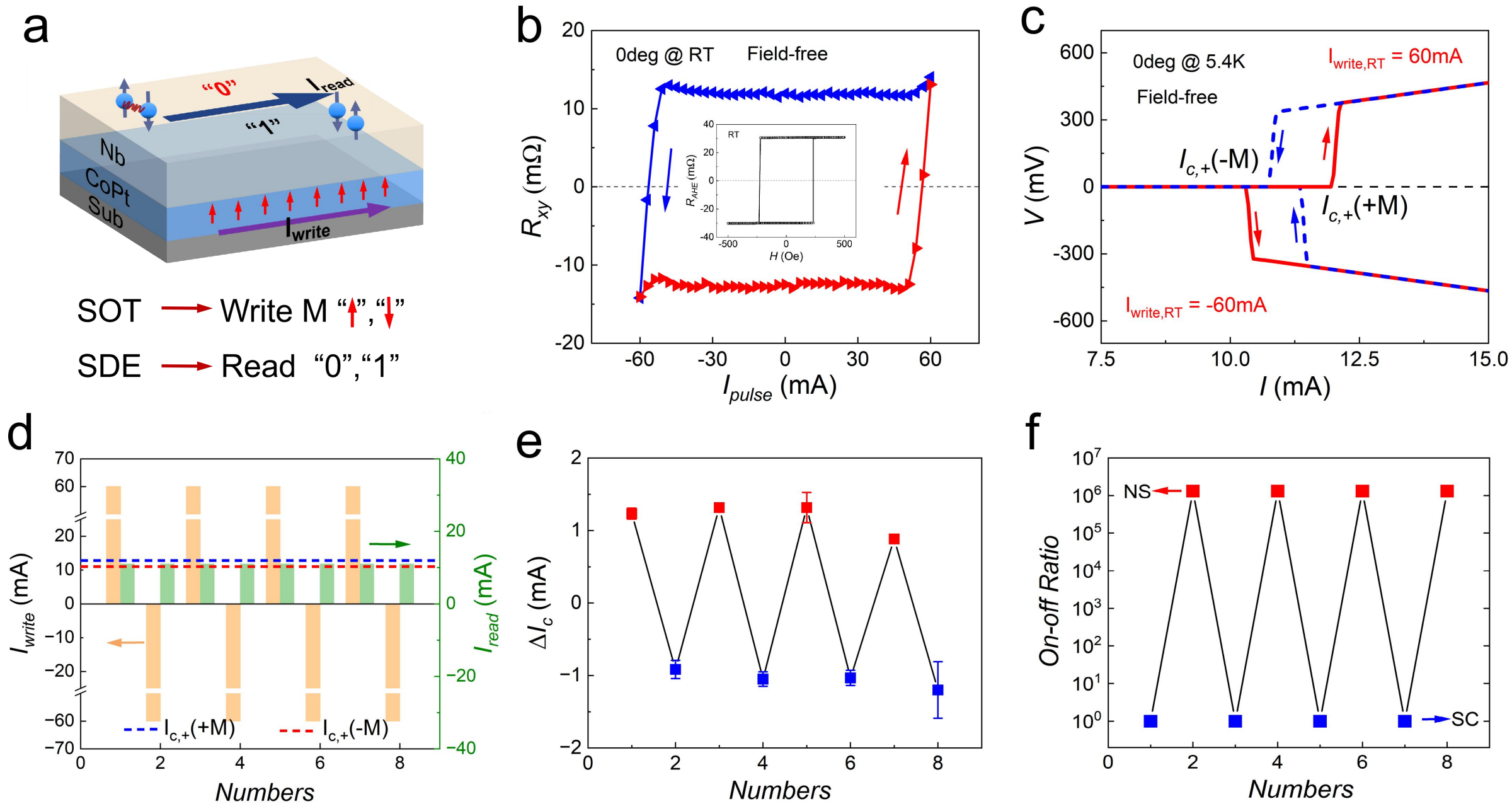

**Figure 5.** Integrated SOT writing and SDE readout in a CoPt/Nb heterostructure. a) Schematic of the write–read architecture based on the CoPt/Nb heterostructure. b) Current-induced magnetization switching of a $\theta$=0° CoPt/Nb device at 300 K and zero-field. The pulse current is swept from $-60$ mA to $+60$ mA and back to $-60$ mA with a step of $2$ mA. The inset shows the anomalous Hall hysteresis loop at $300$ K, confirming robust PMA in the heterostructure. c) Field-free $I - V$ characteristics measured at 5.4 K after writing with $+60$ mA and $-60$ mA pulses, respectively. For clarity, data in the $7.5 - 15$ mA range are compared using identical processing. d) Schematic of repeated write–read cycles. The left axis denotes the room-temperature writing pulse current ($\pm 60$ mA), and the right axis the low-temperature read current (11 mA). The blue and red dashed lines indicate $I_C^+(+M)$ and $I_C^+(-M)$, respectively, extracted from c). e) Extracted $\Delta I_C$ from repeated write–read cycles following the protocol in d). Each point is obtained from three independent $\mathrm{I} - \mathrm{V}$ measurements, with error bars representing the standard deviation. f) Normalized output resistance distribution obtained using $2$ ms pulsed readout under the protocol in d). The two magnetic states correspond to clearly separated normal-state (NS) and superconducting (SC) outputs, demonstrating reliable binary readout functionality.

We then used positive and negative current pulses at 300 K to write the device into the $+M$ and $-M$ states, respectively, followed by zero-field $I - V$ measurements in the superconducting regime. As shown in **Figure 5c**, the two written states exhibit opposite critical-current nonreciprocity. This behavior is consistent with the SDE polarity obtained after preparing opposite CoPt magnetization states using saturation magnetic fields at $10$ K (**Figure S5b, Supporting Information**), indicating that the current-written magnetic states are retained during cooling and continue to control the SDE. Importantly, the opposite diode polarities produce distinct positive critical currents, which can be expressed by $I_C^+(+M) \neq I_C^+(-M)$. For $I_C^+(-M) < I_{read}^+ < I_C^+(+M)$, a single read current yields a superconducting output for $+M$ and a resistive output for $-M$, enabling direct binary readout. Thus, the two nonvolatile magnetic states can be distinguished directly through superconducting and resistive electrical outputs.

To evaluate the operational stability of the device, we performed repeated write–read cycles consisting of room-temperature magnetic writing followed by low-temperature superconducting readout. As shown in **Figure 5d**, the device was sequentially written by positive ($+60$ mA) and negative ($-60$ mA) current pulses at $300$ K, and the output voltage was then measured using a fixed read current ($+11$ mA) at 5.4 K. **Figure 5e** shows that the polarity of the field-free SDE remains controllable by positive and negative write pulses after repeated thermal cycling. Importantly, unlike conventional magnetoresistive readout schemes, which rely on relatively small resistance changes, the present approach uses a transition between a zero-resistance state and a finite-resistance state, producing a clear binary contrast. As shown in **Figure 5f,** the output resistances of the two magnetic states form two well-separated distributions, with an on/off ratio of $10^6$. These results confirm the stability and repeatability of the all-electrical write–read operation.

This device demonstration establishes an all-electrical operation scheme in which field-free SOT writing and magnetic-state-controlled SDE readout are integrated within CoPt/Nb heterostructures. The CoPt magnetization serves as the nonvolatile information carrier, while the SDE converts the magnetic state into either a superconducting or a dissipative electrical output. This approach provides a highly promising route toward ultralow-power cryogenic memory and superconducting spintronic devices.

## CONCLUSIONS

This work demonstrates that intrinsic crystallographic symmetry can govern field-free superconducting nonreciprocity, providing an alternative to approaches that rely on external magnetic fields, asymmetric device geometries, or complex multilayer engineering. In CoPt/Nb heterostructures, the emergence, suppression, and polarity of the SDE are selected by the crystallographic orientation of A1-CoPt and its magnetic state. The absence of comparable nonreciprocity in Nb control devices excludes a dominant contribution from device geometry or random fabrication asymmetry. The identical threefold symmetry dependence of the field-free SDE and field-free SOT switching points to a common crystallographic origin of the two transport responses. Second-harmonic and magnetization-dependent transport further support a vortex-mediated mechanism in which crystal orientation modulates vortex-entry asymmetry, while the CoPt magnetization sets the direction of vortex motion. These results link crystallographic symmetry to mesoscopic vortex dynamics and establish a materials-level route to field-free, polarity-reversible superconducting diodes.

From an integration perspective, the minimal bilayer architecture and conventional device geometry reduce process complexity, facilitating scalable fabrication, device miniaturization and potential compatibility with CMOS-based cryogenic integration. The present write–read demonstration remains a proof of concept, as electrical writing is performed at room temperature followed by superconducting readout at low temperature. Further progress will require reducing the SOT switching current, enlarging the critical-current readout window and realizing in situ magnetic switching in the superconducting regime. More broadly, this symmetry-based strategy may be extended to a broader class of ferromagnet/superconductor heterostructures in which reduced crystal symmetries, such as $C_{1v}$ and $C_{2v}$, provide additional degrees of freedom for engineering nonreciprocal superconducting transport.

## METHODS

**Sample growth**: A1-CoPt (6 nm) films were epitaxially deposited on single-crystal MgO (111) substrates by DC magnetron co-sputtering in an ultra-high-vacuum system with a base pressure below $4 \times 10^{-8}$ Torr. The sputtering powers of Co and Pt targets were adjusted to obtain a nominal composition of $Co_{30}Pt_{70}$. During deposition, the substrate temperature was maintained at 325 °C to promote epitaxial growth and good PMA. After the growth of CoPt, the Nb superconducting layer was deposited at room temperature with a base pressure below $2 \times 10^{-8}$ Torr. The sputtering power was optimized to ensure good superconducting properties of Nb. The resulting heterostructure was MgO//CoPt(6 nm)/Nb(30 nm). The films were subsequently patterned into Hall bar devices with a channel width of $5\ \mu m$ using photolithography and Ar ion milling.

**Structural characterization**: The crystalline structure of the films was characterized by high-resolution X-ray diffraction (HR-XRD). $\theta - 2\theta$ scans were performed to determine the crystal phase and out-of-plane texture, while $\varphi\ scans$ were used to examine the in-plane epitaxial relationship and crystal symmetry.

**Electrical transport and magnetization measurements**: AHE measurements were performed to characterize the PMA and current-induced magnetization switching of CoPt. The transverse Hall voltage $V_{xy}$ was measured at both room temperature and low temperature using a Keithley 6221 current source and a Keithley 2182A nanovoltmeter with a dc current of $1\ mA$. For the angle-dependent SOT switching measurements of the CoPt films, current pulses with a duration of 30 μs were used. The superconducting-state pulsed readout in Figure 5f was also performed using 2 ms pulses. Superconducting transport measurements were carried out in a Multi-Field ColdTUBE and a Quantum Design PPMS. The SDE was characterized by measuring $I - V$ curves, where the current was supplied by a Keithley 6221 and the voltage was recorded by a Keithley 2182A nanovoltmeter. First- and second-harmonic transport measurements were performed using two Stanford Research Systems SR830 lock-in amplifiers, enabling quantitative analysis of the nonreciprocal transport coefficient in CoPt/Nb heterostructures.